\documentclass[secnumarabi,superscriptaddress,amssymb,nofootinbib,aps,prc]{revtex4}

\usepackage{graphicx}
\usepackage{amsmath}
\usepackage{hyperref}
\usepackage{bm}
\usepackage{color}

\begin{document}

\title{Exploring QCD dynamics in medium energy  $\gamma  A$ semiexclusive  collisions}

\author{A.B. Larionov}
\affiliation{Institut f\"ur Theoretische Physik, Universit\"at Giessen,
             D-35392 Giessen, Germany}
\affiliation{National Research Centre "Kurchatov Institute", 
             123182 Moscow, Russia}
\author{M. Strikman}
\affiliation{Department of Physics, the Pennsylvania State University, State College, PA 16802, USA}
\date{\today}

\begin{abstract}
We demonstrate that studies of the semiexclusive large angle photon - nucleus  reactions: $\gamma + A\to h_1+h_2 +(A-1)^*$ with tagged photon beams of energies $6 \div 10$ GeV 
which can be performed in Hall D at Thomas Jefferson National Acceleration Facility (TJNAF) would allow to probe several aspects of the QCD dynamics:
establish the $t$-range in which transition from soft to hard dynamics occurs, compare the strength of the interaction of various mesons and baryons with nucleons
at the energies of few GeV, as well as look for the color transparency effects. 
\end{abstract}
\maketitle 

\vspace{3mm}

 Keywords:  $\gamma A$ reactions,  QCD

\vspace{3mm}
PACS: 12.38.-t,   24.85.+p
\vspace{3mm}

\section{Introduction}
\label{Intro}
Recently an extensive program of experiments with tagged photon beams of energies up to about 10 GeV  has started at TJNAF in the Hall D. 
The prime aim of these experiments is to explore the QCD dynamics of mesons using polarized photon beams and proton targets. 
The high photon fluxes, photon tagging and good acceptance of the detector allow also to study various rare exclusive processes 
such as large angle photon-nucleon scattering with production of a meson and a baryon.

It is presently a well accepted concept that the hadron fluctuates between various quark-gluon configurations since it is a quantum-mechanical
system of quarks and gluons.
Various processes involving the hadron test various configurations. For example in $t$-channel processes, the wave length of the exchanged
particle should match the size of a configuration. Hence the hard binary processes, where not only the center-of-mass (c.m.) energy, $\sqrt{s}$,
is large, but also $|t| \sim |u| \sim |s|$, select the small-size configurations in the hadron. We will call that selection ``squeezing'' below and the
selected color-neutral configurations -- ``squeezed configurations''. An extreme case of linear size much smaller than the average size
is reached for the point-like configurations  (plc's).

We will demonstrate here that exclusive hard two-body large c.m. angle processes with nuclear targets:
\begin{equation}
  \gamma + A \to h_1 +h_2 + (A-1)^*~,
\label{aqu}
\end{equation}
would allow to address a number of important aspects of the QCD dynamics.
We will focus on three questions:
(a) At what momentum transfers, $t$, does the transition from the regime in which the photon interacts as a vector meson 
to the one dominated by the contribution of unresolved (i.e. point-like) photon occur ?
(b) How different are the interactions of different mesons (baryons) at energies of few GeV ? 
(c) At what kinematics are the outgoing hadrons produced in the squeezed configurations and weakly 
interacting with other nucleons on the way out of the nucleus due to the color transparency (CT) ?
Note that we often refer to the squeezed configurations rather than to the plc's to emphasize that CT effects are likely to set in already
when the linear size of the quark configuration is just  $30\div40\%$ smaller than the average size.
Comprehensive reviews of the CT can be found in refs. \cite{Frankfurt:1994hf,Jain:1995dd,Dutta:2012ii}.
The reactions of Eq.(\ref{aqu}) also represent one of the best tools for study of the short-range correlations and non-nucleonic degrees of freedom 
in nuclei \cite{Farrar:1988mf,Piasetzky:2006ai}.

Understanding the dynamics of two-body large momentum processes remains a challenge for the QCD for a long time.
In these processes, large momenta are transferred to the outgoing hadrons, so one may naively expect that the process should be described by the perturbative QCD.
Indeed, the quark counting rules \cite{Brodsky:1973kr} based on the consideration of the lowest order QCD diagrams involving all constituents
of the initial and final hadrons describe the energy dependence of the cross section at fixed c.m. scattering angle, $\Theta_{c.m.}$,
pretty well (though the data are rather limited) both for hadronic \cite{White:1994tj} and photon \cite{Anderson:1976ph} projectiles:
  \begin{equation}
  {d\sigma \over d \Omega_{c.m.}} = {1 \over s^{\sum n_i -2}} f(\cos\Theta_{c.m.})~, 
  \label{counting}
  \end{equation}
where $n_i$ is the number of the constituents in the incoming and outgoing  particles:
  \begin{equation}
     n_B=3,~n_M =2,~n_\gamma=1~.
  \label{n_const}
  \end{equation}
  A notable exception is the large angle Compton scattering for which the cross section is found to drop as $s^{-7}$ 
rather than $s^{-6}$ \cite{Danagoulian:2007gs}.

So far no successful calculations of the absolute cross sections were reported for the large c.m. angle scattering 
processes. The basic assumption of the quark counting rules is that the incoming and outgoing particles in the interaction point are
in configurations with a much smaller than the average size -- the plc's: $r_t \propto 1/\sqrt{-t}$. 
If a hadron propagates a long distance in a plc, the strength of the absorption would be strongly suppressed 
due to the color transparency property of the QCD -- interaction of a color singlet small transverse size configuration
is proportional to the area occupied by the color. Hence, it was suggested to check the presence of squeezing 
in the large angle processes  by studying their $A$-dependence \cite{Mueller82, Brodsky82}.
In the limit of strong squeezing, the regime of full transparency was predicted:
  \begin{equation}
  T(A) = {d\sigma(a + A  \to b +N + (A-1)^* ) \over d\Omega_{c.m.}}/
  {Zd\sigma(a + p  \to b +N) \over d\Omega_{c.m.}}=1~,
  \end{equation}
where we considered the process in which the elementary reaction involves scattering off protons 
\footnote{For simplicity, we neglect the Fermi motion effects here and below.}.
However, it was demonstrated  in \cite{Farrar:1988me} that the space-time evolution of the hadron wave packets 
involved in the hard interaction strongly reduces the transparency in a wide range of energies since even in the case 
when hadrons are squeezed to plc's in the point of hard interaction - these configurations are not frozen 
and expand increasing the chances of absorption. 
   
The CT was observed in high energy coherent diffraction of pion into two jets \cite{Aitala:2000hc} in line with the predictions of \cite{Frankfurt:1993it}.
The measurements of the transparency in the electroproduction of pions and $\rho$-mesons at TJNAF are consistent with the prediction of the models which include squeezing
and quantum diffusion \cite{Farrar:1988me}. At the same time the origin of the energy dependence  of $T$ in the reaction $pA\to pp(A-1)^*$ remains unclear
and there seems to be no evidence for the CT effects in the $(e,e'p)$ reaction in the $Q^2$ range up to 8 GeV$^2$ studied so far \cite{Dutta:2012ii}.

This  suggests that the hard exclusive  reactions in which mesons are involved are more promising for observing CT effects than other exclusive processes. 
Photon beams provide an additional advantage as one can explore at what momentum transfers the transition from the dominance of vector-meson-like
configurations (resolved photon) to the dominance of unresolved photons  occurs -- this transition may be referred to as the onset of 
the "photon transparency". 
Another advantage of the photon beams is variety of final states: {\it Mesons} = $\pi, \eta, \rho...$, {\it Baryons}= $N, \Delta$ 
which would allow to compare the properties of different hadrons.

Note here that the squeezing is expected also for a novel class of hard branching hadronic processes $a + b \to c +d +e$, where the hadrons $c$ and $d$ have large
and nearly opposite transverse momenta and large invariant energy which is a finite fraction of the total invariant energy \cite{Kumano:2009he}.
An interesting example of such a reaction is the production of two mesons (or a meson and a photon) in the photon - nucleus scattering \cite{Beiyad:2010cxa,Boussarie:2016aoq}.
Color transparency effects are enhanced for these reactions as compared to two-body reactions \cite{Kumano:2009ky}.

The paper is organized as follows. In section \ref{formalism} we summarize the basic formulae necessary for the calculation of the nuclear transparency.
In section \ref{photoTran} we consider the photon transparency. It is shown that $T$ increases very significantly in the regime of photon transparency.
In section \ref{inelCS} we demonstrate that the discussed process with different final states would allow to perform a critical test of the constituent quark model 
by comparing the cross sections of inelastic interactions of different mesons - say pion and $\eta$ mesons. The estimates of the color transparency effect 
for the highest energies available at TJNAF are presented in section \ref{CT}. Our findings are summarized in section \ref{concl}.
     
     \section{Basic formulae}
      \label{formalism} 
    
We will consider the particular case of Eq.(\ref{aqu}) when one of the outgoing particles is a nucleon. 
For the discussed reactions $\gamma + A \to \text{h}^\prime + \text{N}^\prime + (A-1)^*$ 
with an outgoing hadron $\text{h}^\prime$ and a nucleon $\text{N}^\prime$, the transparency
can be estimated in a semiclassical approximation taking into account the absorption
due to the inelastic interactions of incoming and outgoing 
particles\footnote{Calculations of the transparency in the Glauber approach within the 
Hartree-Fock-Skyrme model of the nucleus give very similar results \cite{Frankfurt:1994nn}.}:
 \begin{eqnarray} \label{eq6.20}
   T & = & {N_q}^{-1} \int \text{d}^{2}b \; \text{d}z
     \; \rho_q(b, z) \exp \biggl( - \sigma^{\text{eff}}_{\gamma\text{N}} \int\limits^{z}_{z-l_\gamma}
     \text{d}z^{\prime} \; \rho(b,z^{\prime}) \nonumber \\
   & - & \int\limits^{\infty}_{l_r} \text{d}l \; \rho(b_r,l)
     \sigma_{\text{h}^{\prime} \text{N}}^{\text{eff}}(p_{\text{h}^{\prime}},l-l_r) -
   \int\limits^{\infty}_{l_r^\prime} \text{d}l^{\prime} \; \rho(b_{r^\prime},l^\prime) 
   \sigma_{\text{N}^\prime\text{N}}^{\text{eff}}(p_N^\prime,l^\prime-l_r^\prime) \biggl).
\end{eqnarray}
Here $z$ is the coordinate along the direction of the photon momentum, $b$ is the impact parameter of the photon. 
$\rho_q(\mathbf{r}),~q=p,n$ represents the density of protons or neutrons, depending on which nucleon the hard process happens. 
The densities are normalized to the corresponding particle numbers: $\int d^3 r\rho_q(\mathbf{r}) =N_q$,
where $N_p \equiv Z$ and $N_n \equiv A-Z$. $\rho(\mathbf{r})=\rho_p(\mathbf{r})+\rho_n(\mathbf{r})$ is the total nucleon density.
The quantities $\sigma^{\text{eff}}$ are the effective absorption cross sections of the incoming and outgoing particles which
may generally depend on the distance from the hard interaction point (see below). 
$l$ and $l^{\prime}$ denote the coordinates along the linear trajectories of the outgoing hadron $\text{h}^\prime$ and nucleon $\text{N}^\prime$, respectively.
The corresponding initial values and impact parameters are determined by assuming the spherical symmetry of the target nucleus which gives:
$l_r=\mathbf{r}\mathbf{p}_{\text{h}^\prime} /p_{\text{h}^\prime}$, $b_r=\sqrt{r^2-l_r^2}$, 
$l_r^\prime=\mathbf{r}\mathbf{p}_{\text{N}^\prime}/p_{\text{N}^\prime}$, $b_{r^\prime}=\sqrt{r^2-(l_r^\prime)^2}$,
where $\mathbf{r} \equiv (\mathbf{b},z)$. Note that due to the spherical symmetry, the r.h.s of Eq.(\ref{eq6.20}) does not depend on the azimuthal angles 
of the outgoing particles (which differ by $\pi$), while their longitudinal and transverse momenta are fully determined by the kinematic invariants  
$t=(p_{\text{h}^\prime}-p_\gamma)^2$ and $s=(E_\gamma+E_N)^2-p_{\rm lab}^2$.
Here $E_\gamma=p_{\rm lab}$, and $E_N=m_N-B$ where $B \simeq 40$ MeV is the removal energy of the nucleon at rest
which includes the average binding energy of the nucleon ($\sim 8$ MeV) and the excitation energy of the residual nucleus ($\sim E_{\rm F}$).
Other choice of the nucleon removal energy, e.g. setting $B=0$, practically does not influence the numerical results below,
except the region of very large $-t$ (backward scattering) which is not the main focus of this study.  

We will deal here with photon-nucleus reactions at $p_{\rm lab} \sim 5\div20$ GeV/c where the photon coherence length,
\begin{equation}
    l_\gamma=\frac{2p_{\rm lab}}{m_\rho^2},   \label{l_gamma}
\end{equation}
is comparable or larger than the medium-mass nucleus radius $\sim 4$ fm. This means that the photon coherently transforms to the $\rho$ meson which propagates 
the distance up to $\sim l_\gamma$ inside the nucleus before it experiences incoherent interaction with a nucleon producing the final hadron $\text{h}^\prime$.
As well known \cite{Bauer:1977iq}, the interference of 
this two-step process with the one-step process (i.e. when the photon incoherently produces $\text{h}^\prime$
without any coherent interactions before) leads to the modification of the effective mass number apart from the usual absorption factor for 
an outgoing
particle. In rough approximation, this effect is expressed in Eq.(\ref{eq6.20}) as an absorption factor for the incoming photon which interacts with nucleons 
with effective cross section $\sigma^{\text{eff}}_{\gamma\text{N}}$. However, this additional absorption is well established for the diffractive processes only, 
when the transverse momentum transfer is small. It is not clear a priori whether the photon 
behaves as a point-like particle or as a $q\bar q$ pair in the hard photoproduction processes studied here. 
Thus we perform calculations in two regimes: In the ``resolved photon'' (RP) regime, 
the effective photon absorption cross section $\sigma^{\text{eff}}_{\gamma\text{N}} = \sigma_{\rm in}(\pi N)$ is applied,
where $\sigma_{\rm in}(\pi N)$ is the inelastic pion-nucleon cross section.
In the ``unresolved photon'' (UP) regime, we set $\sigma^{\text{eff}}_{\gamma\text{N}}=0$. In default case the UP regime is assumed.
   
The CT effects are accounted for within the quantum diffusion model \cite{Farrar:1988me} by using the following effective cross sections 
in Eq.(\ref{eq6.20}):
\begin{equation}
   \sigma_{h N}^{\rm eff}(p_h,z)
  =\sigma_{h N}(p_h) \left(\left[ \frac{z}{l_h}
    + \frac{\langle n_h^2k_t^2\rangle}{M_{\rm CT}^2} \left(1-\frac{z}{l_h}\right) \right]
    \Theta(l_h-z) +\Theta(z-l_h)\right)~,~~~~h=h^\prime,~N^\prime~,         \label{sigma^eff}
\end{equation}
where $z$ is the distance from the hard interaction point and $M_{\rm CT}^2=\min(-t,-u)$ is the CT scale.
In the spirit of the semiclassical approximation which we apply here,  $\sigma_{h N}(p_h)$ is set equal to the inelastic hadron-nucleon cross section
\footnote{In the Glauber theory $\sigma_{h N}(p_h)$ should be set equal to the total hadron-nucleon cross section. However, elastic rescattering
terms neglected in Eq.(\ref{eq6.20}) should effectively increase the transparency at large momentum transfer. This can be approximately treated by 
using the inelastic cross sections in the exponential absorption factors.}~.
Other quantities in Eq.(\ref{sigma^eff}) are defined as follows: $n_h$ is the number of constituents (Eq.(\ref{n_const})),
$\langle k_t^2 \rangle^{1/2} \simeq 0.35$ GeV/c is the average transverse momentum of a quark in a hadron, and
\begin{equation}
    l_h=\frac{2p_h}{\Delta M^2}                   \label{l_h}
\end{equation}
is the hadron coherence length. The mass denominator $\Delta M^2$ represents the major uncertainty of the quantum diffusion model.
To reduce the number of free parameters, we assume the same $\Delta M^2$ for a pion and a proton. The calculations are performed
with the two values, $\Delta M^2=0.7$ GeV$^2$ and 1.1 GeV$^2$.  The value of 0.7 GeV$^2$ is favored by the analyses \cite{Larson:2006ge, Larionov:2016phv}
of the nuclear transparency in $A(e,e^\prime \pi^+)$ reaction in the collinear kinematics measured at TJNAF \cite{Clasie:2007aa}.
The value of 1.1 GeV$^2$ is close to the difference of masses squared of a $N(1440)$ resonance and a nucleon.

Below, the calculation with vanishing hadron coherence length, $l_h=0$, will be referred to as a ``Glauber'' one, while the calculation with
$l_h$ of Eq.(\ref{l_h}) -- as a ``quantum diffusion'' or ``CT'' one.
All calculations are performed in the kinematics of the process $\gamma + A \to \pi^- + p + (A-1)^*$.

    \section{Photon transparency effect}
      \label{photoTran} 

In soft intermediate energy hadroproduction processes the photon is usually treated as a superposition of hadronic states of moderate masses: 
the lightest vector mesons and higher mass states. The situation is expected to be different in the large angle exclusive production at sufficiently
large energies since the cross section of the process in which the photon is resolved is suppressed by a factor $1/s$ as compared to the contribution
of the unresolved photon -- cf. Eq.(\ref{counting})  with $n_\gamma=2 $ and $n_\gamma =1$. 
It was found in \cite{Anderson:1976ph} that the $s^{-7}$ behavior of $d\sigma/dt$ at a fixed angle represents the data over a wide range of angles 
for several exclusive processes $\gamma p \to \mbox{meson}+\mbox{baryon}$ studied there. Note, however, that the data do not exclude a somewhat
higher power of $1/s$. The behavior consistent with the unresolved photon contribution appears to set in at  $-t \gtrsim 2~\mbox{GeV}^2$.   
  
\begin{figure}
\begin{center}
   \includegraphics[scale = 0.7]{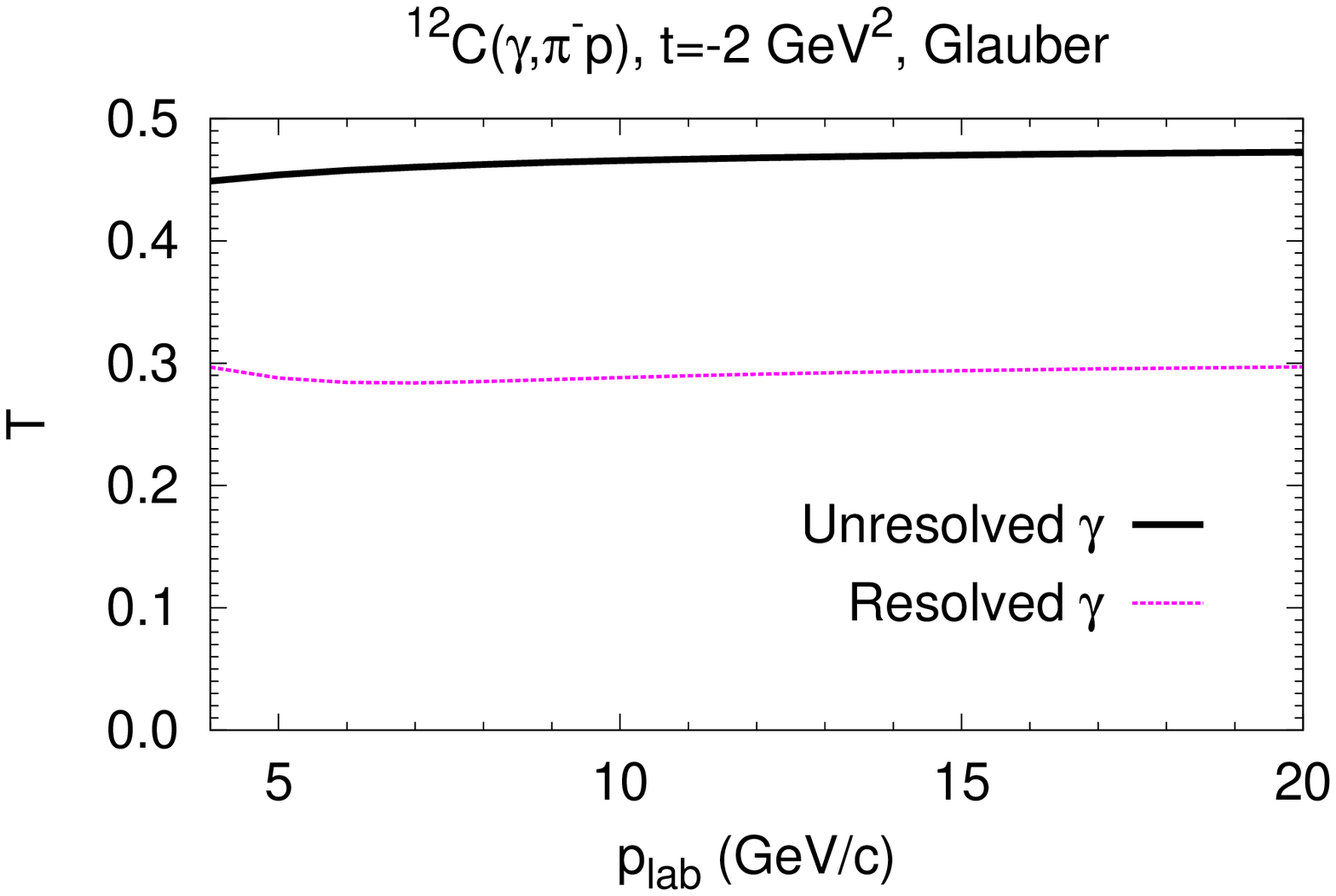}
   \includegraphics[scale = 0.7]{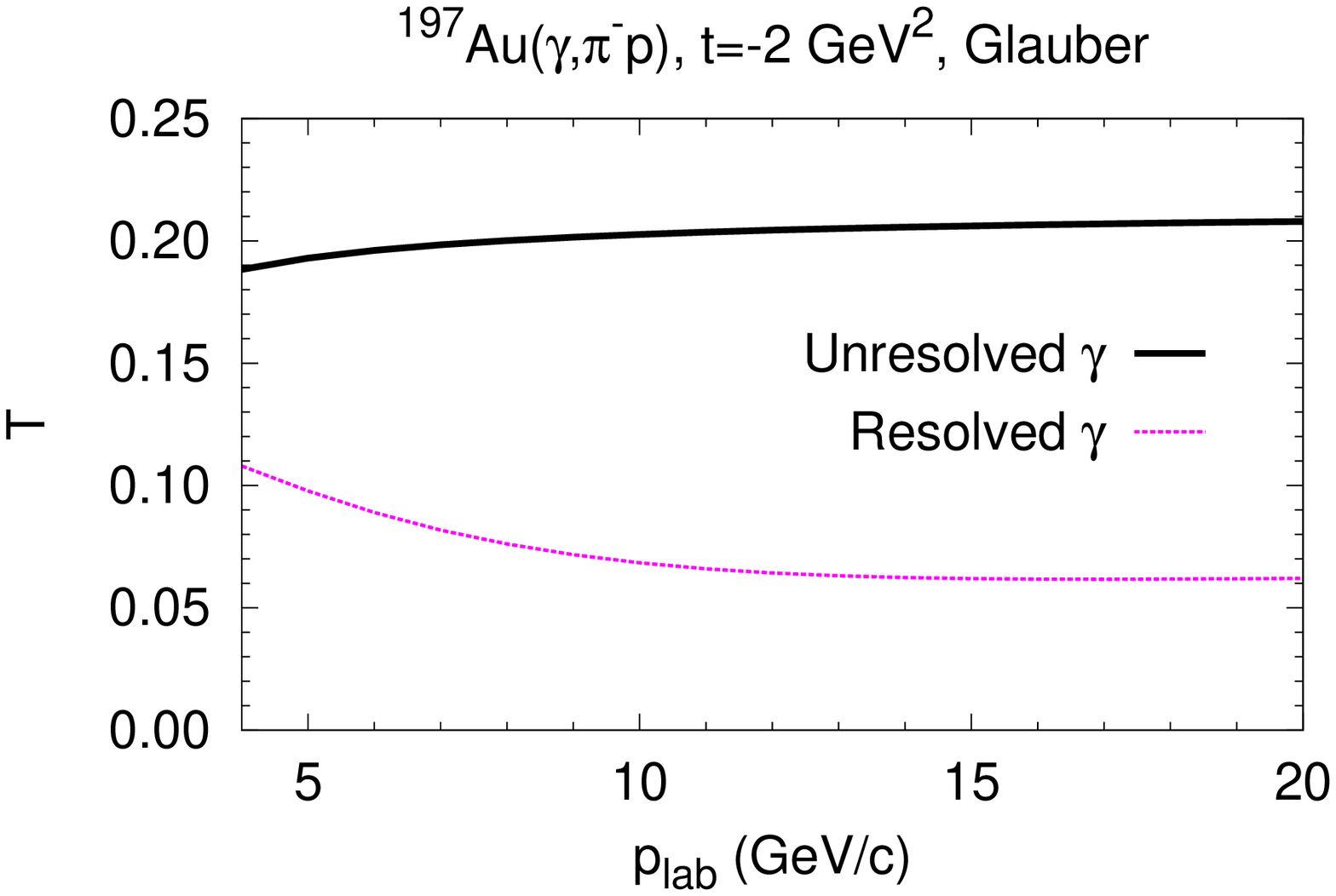}
\end{center}
\caption{\label{fig:photran_t-2gev2} Transparency for the $A(\gamma,\pi^-p)$ semiexclusive process for $^{12}$C and $^{197}$Au target nuclei   
under assumption  of the dominance of the unresolved photon ($\sigma^{\text{eff}}_{\gamma\text{N}}=0$, solid curve) and resolved photon interacting in configuration with 
$\sigma^{\text{eff}}_{\gamma\text{N}}=\sigma_{\rm in}(\pi N)$ (dashed curve). Calculations are performed at $t=-2$ GeV$^2$.}
\end{figure}

\begin{figure}
\begin{center}
   \includegraphics[scale = 0.7]{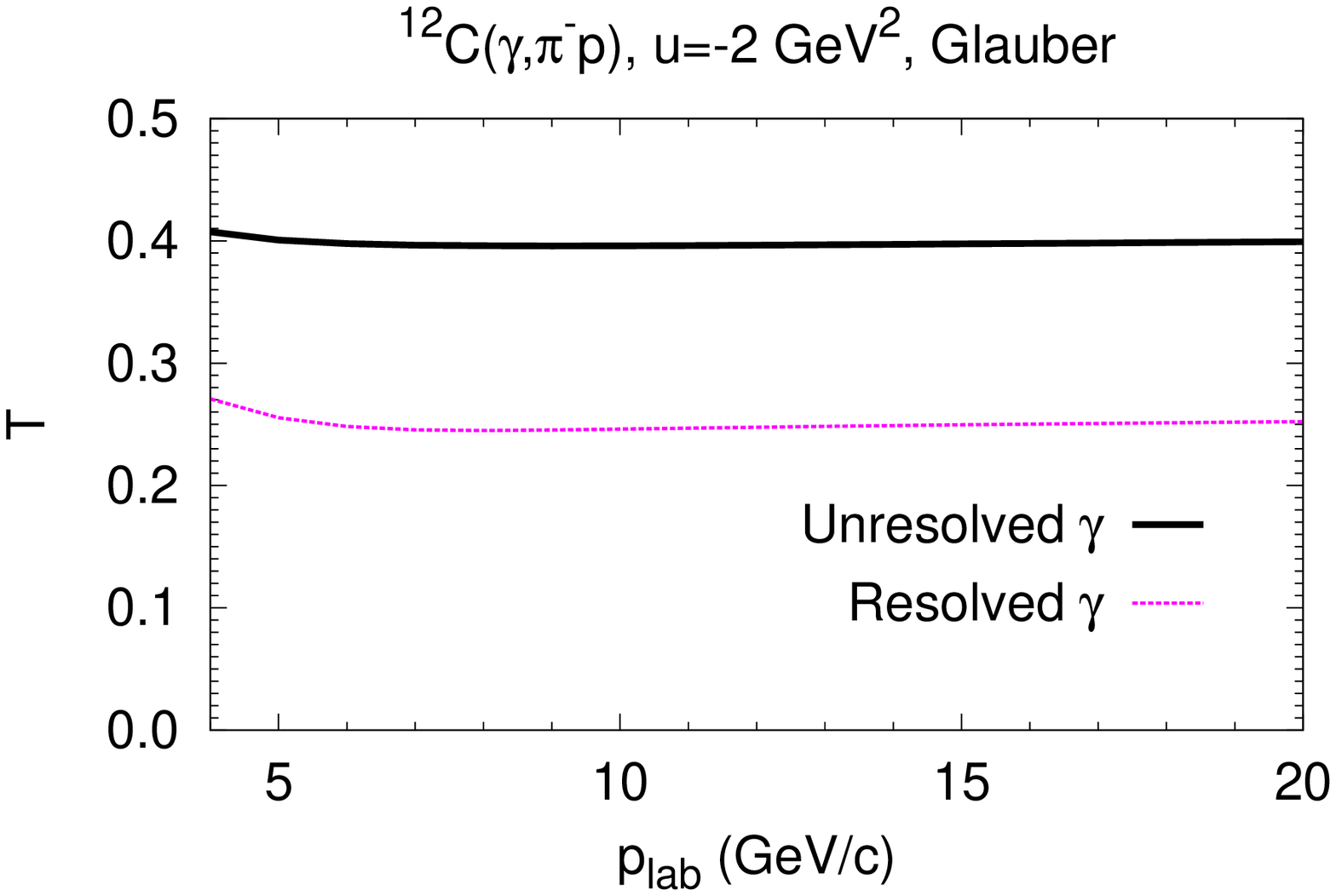}
   \includegraphics[scale = 0.7]{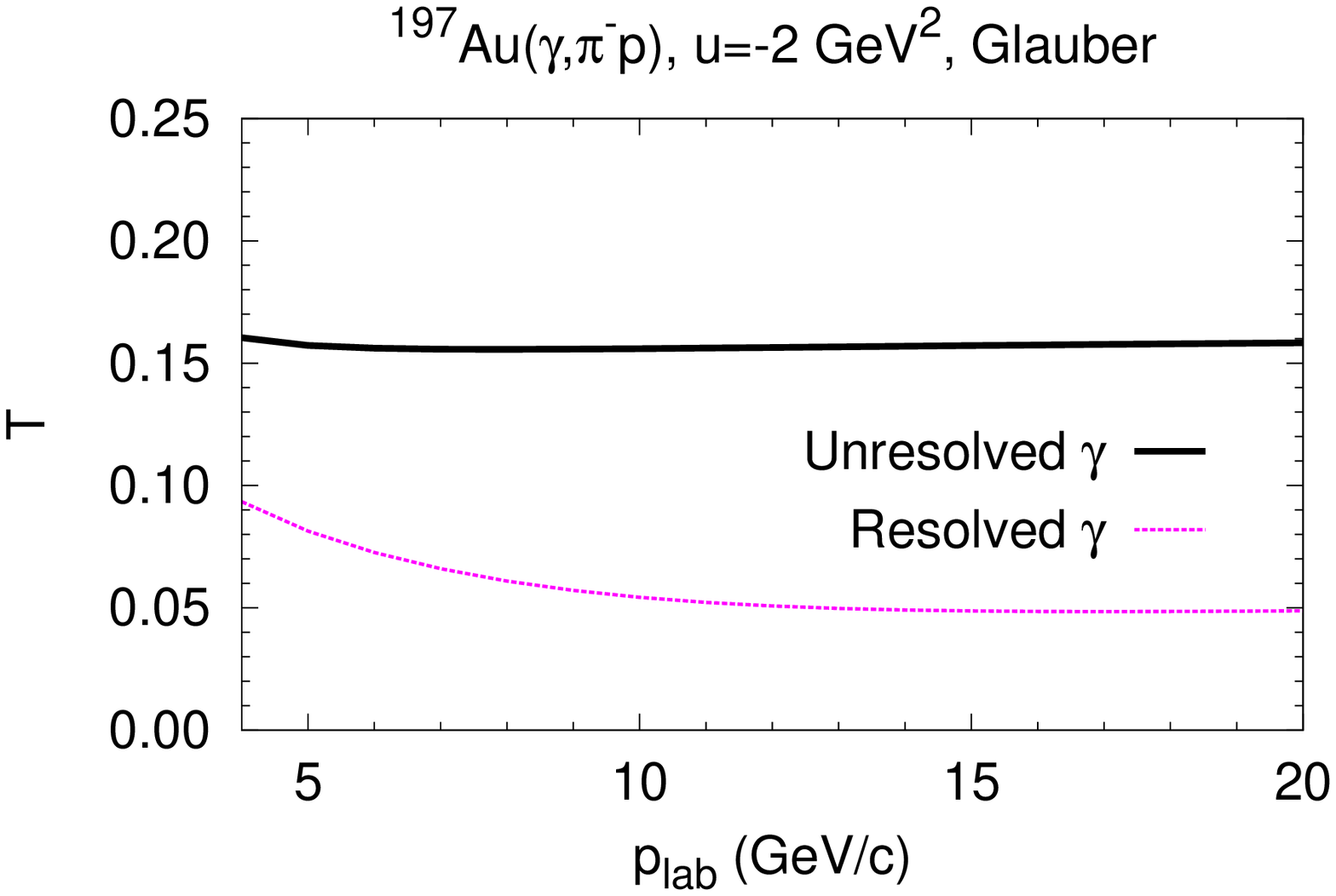}
\end{center}
\caption{\label{fig:photran_u-2gev2} Same as Fig.~\ref{fig:photran_t-2gev2}, but for fixed $-u=2$ GeV$^2$.}
\end{figure}

  Scattering off nuclei provides a sensitive probe of the transition from the RP to the UP regime, as displayed 
in Figs.~\ref{fig:photran_t-2gev2},\ref{fig:photran_u-2gev2}.
In the RP regime the photon coherence length, $l_\gamma$, grows with increasing beam momentum, and the photon absorption region extends over large 
distances ultimately exceeding the nucleus diameter.
This results in the moderate decrease and saturation of the transparency with growing $p_{\rm lab}$.
However, in the UP regime the photon penetrates the nucleus without 
interaction up to  the hard interaction point (probability of its transition to a hadronic configuration during passing through the nucleus is suppressed by 
the small value of the fine-structure constant, $\alpha_{e.m.}$). Thus, in this case the photon can interact with all nucleons near the back surface 
of the nucleus producing a pair of hadrons which can escape without absorption, leading to  $\sigma \propto A^{2/3}$. 
In contrast, in the resolved photon (RP) regime only the rim of the nucleus contributes into $T$ ($\sigma \propto A^{1/3}$). 
As a result, under the UP scenario the transparency is much higher than in the RP scenario in which the photon interacts in a hadronic configuration 
\footnote{As far as we know this effect was first briefly mentioned in  \cite{Miller:2010eh}. 
It was also discussed in the talks of the authors at TJNAF.}.  
Moreover, the ratio of transparencies predicted by the UP and RP models is expected to grow strongly with $A$ ($\propto A^{1/3}$).
Of course, the applicability of these qualitative estimates depends on the values of cross sections, nuclear density profiles etc.
Our numerical calculations of the nuclear transparency at $p_{\rm lab}=10$ GeV/c, $t=-2$ GeV$^2$ for the different nuclear targets with masses
numbers between 12 and 208 can be well fit with the power law,
\begin{equation}
  T=A^{\alpha-1}~,     \label{PowerLaw}
\end{equation}
where $\alpha=0.704\pm0.002$ for the UP regime and $\alpha=0.509\pm0.003$ in the RP regime.
Thus, the above estimate ($\alpha=1/3$)  should be taken as a lower limit for the nuclear transparency for the RP regime.

The UP and RP regimes strongly differ even at relatively small beam momenta, $\sim 5$ GeV/c,
since the photon coherence length $l_\gamma$, Eq.(\ref{l_gamma}), is large and
the photon attenuation has a full strength within the range $l_\gamma$ before the hard interaction point.
This is qualitatively different from the pattern  predicted in the color transparency regime 
(cf. Fig.~\ref{fig:photran_t-2gev2} and Figs.~\ref{fig:T_90deg}-\ref{fig:T_t})  
due to the hadron coherence length, $l_h$ (Eq.(\ref{l_h})), being
much smaller than the radii of heavy nuclei for the discussed photon energy range
(since the beam momentum is shared between the two produced hadrons).

Hence, we conclude that it would be feasible to study the transition to the photon transparency regime at TJNAF. 
It would be important to perform such studies for different channels - since the onset of this regime may depend on the quark composition 
($\pi $ vs $\eta$ mesons) and the spin of the final hadrons ($\pi $ vs $\rho$ mesons).

As we see from Fig.~\ref{fig:photran_u-2gev2}, in the kinematics with fixed $-u = 2~\mbox{GeV}^2$, the corresponding transparencies 
are $\sim 20\%$ smaller than those at fixed $-t = \, 2~\mbox{GeV}^2$. This is because in the former case the fast forward proton has larger 
inelastic cross section.
The power law fit, Eq.(\ref{PowerLaw}), of the mass dependence at $p_{\rm lab}=10$ GeV/c, $-u = 2~\mbox{GeV}^2$
produces the values $\alpha=0.650\pm0.003$ for the UP regime and $\alpha=0.457\pm0.004$ in the RP regime.
The change of $T$ due to the photon transparency is similar for the kinematics with fixed $t$ and $u$.
However, since the onset of the photon transparency regime may start at different values of the momentum transfer in $t$- and $u$- channels,
it would be desirable to study experimentally both $-t \sim \, 2~\mbox{GeV}^2$ and $-u \sim \, 2~\mbox{GeV}^2$ kinematics.

    \section{Comparing inelastic cross sections of meson - nucleon interactions }
          \label{inelCS} 

  It is well known that the additive quark model describes well the relations between cross sections of $\pi N, K N$ and $NN$ scattering \cite{Levin:1965mi}. 
However, there are indications that such relations should break for scattering of other mesons.
For example, let us consider the cross section of $\eta N$ interaction. In the quark model  (neglecting small non ideal mixing)
\[
   \left |\eta \right > = {\left |u\bar u \right > +\left |d\bar d \right > - 2 \left |s\bar s \right >\over \sqrt{6}}~.
\]
Hence, in the additive quark model
\begin{equation}
\sigma(\eta N) = \frac{\sigma(u\bar u N) + 2\sigma(s\bar s N)}{3}      
 = \frac{4 \sigma (K N)  - \sigma (\pi N)}{3} \sim 16~\text{mb}~,
\end{equation}
where we took $\sigma(qN)= \sigma(\bar q N),~\sigma(uN)=\sigma(dN)$. 
This is much smaller than $\sigma_{\rm in}(\eta N)  \approx$ 30 mb extracted from the measurements of the $A$-dependence of cross section of photon-nucleus 
interactions \cite{Mertens:2008np} at $E_\gamma \leq 2.2$ GeV. Similarly, the measurements of $\phi$-meson absorption in 
photon-induced reactions at $E_\gamma=1.5\div2.4$ GeV reported the values of $\sigma_{in}(\phi N) \sim $ 35 mb \cite{Ishikawa:2004id} which are much larger than 
$\sigma_{\rm in}(\phi N) \sim$ 10 mb expected in the additive quark model.
  
\begin{figure}
\begin{center}
   \includegraphics[scale = 0.6]{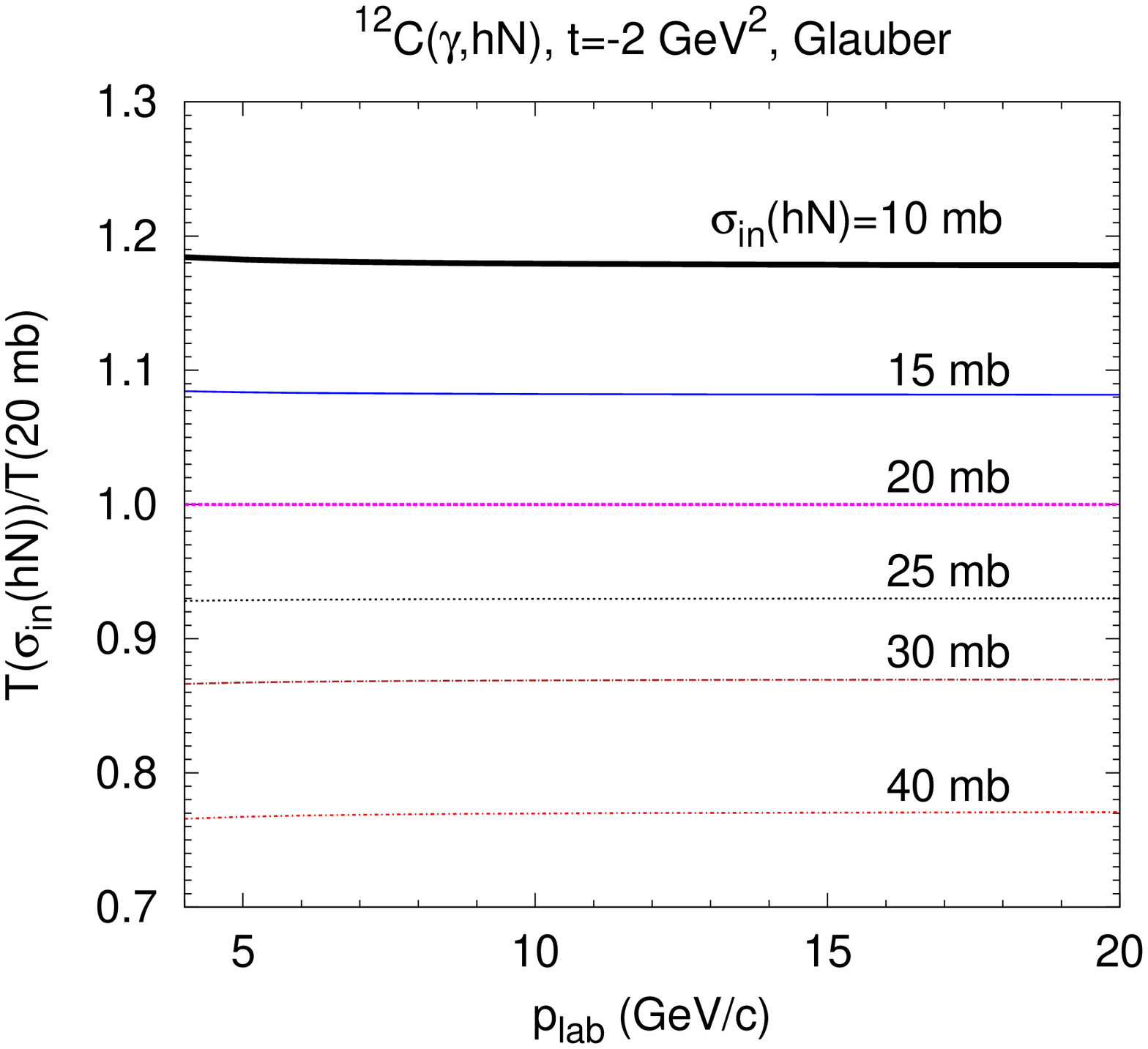}
   \includegraphics[scale = 0.6]{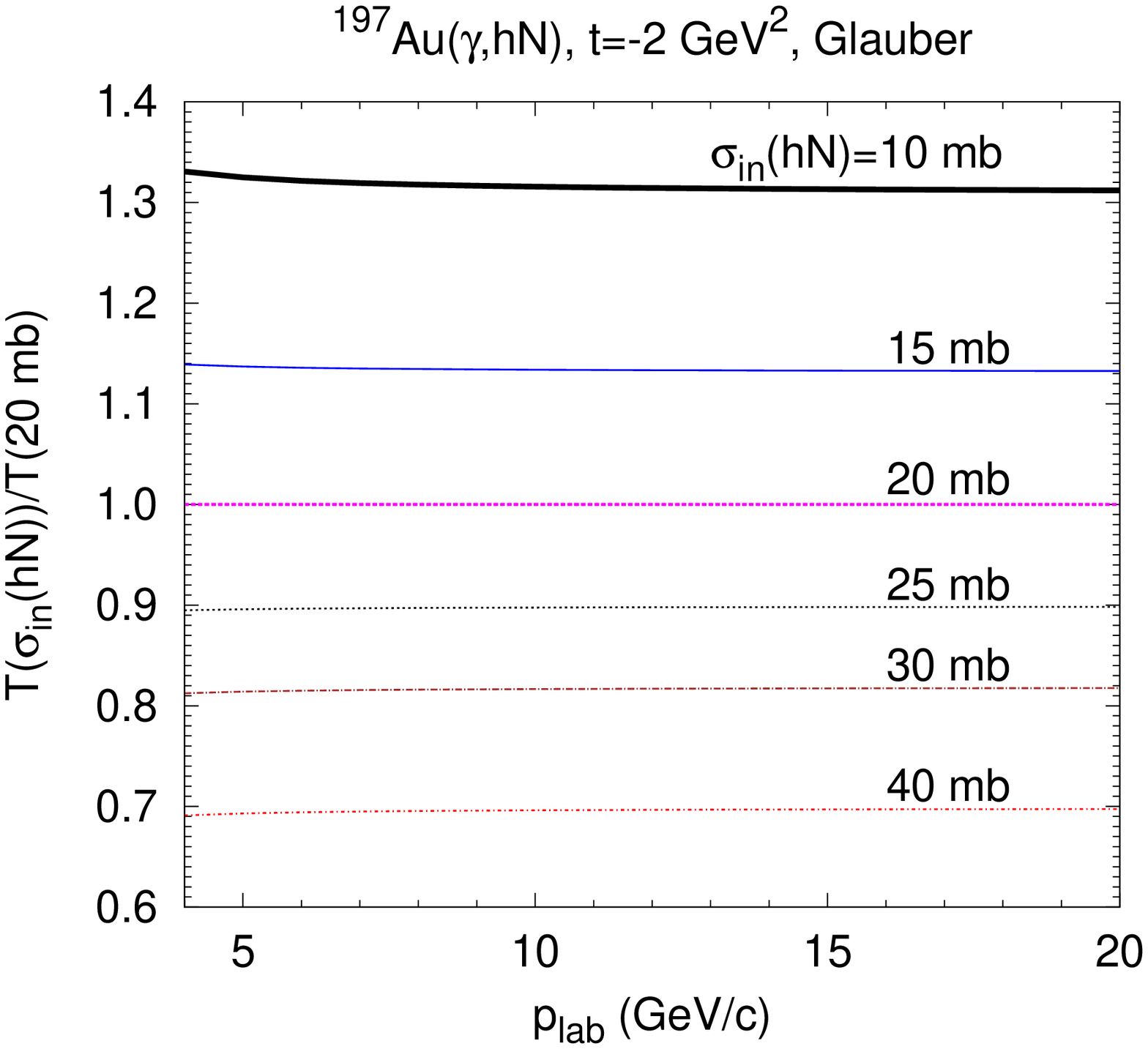}
\end{center}
\caption{\label{teta} Dependence of the transparency in the reaction 
$\gamma A \to h N (A-1)^*$ on $\sigma_{\rm in}(hN)$ normalized to the transparency for $\sigma_{\rm in}(\pi N)$=20 mb,
where $\sigma_{\rm in}(\pi N)$ is the inelastic $\pi^+ p$ cross section at $p_{\rm lab} \sim 10\div100$ GeV/c.}
\end{figure}

 To quantify the strength of absorption in the discussed class of reactions one can measure  the ratio of the transparencies for, say, 
$\phi$- or $\eta$-meson production and for the reference process $\gamma A \to \pi N$ for the same kinematics. 
For example, one can greatly reduce nuclear structure effects by considering  the double ratio
 \begin{equation}
 R(A) = {\sigma(\gamma +A\to h_1+N + (A-1)^*)\over \sigma(\gamma +A\to h_2+N + (A-1)^*)}
/ {\sigma(\gamma +N\to h_1+N)\over \sigma(\gamma +N\to h_2+N)}~.
\end{equation}
The results of our numerical calculations under assumption that the photon transparency (i.e. UP) regime sets in already at $-t=\mbox{2  GeV}^2$ 
are presented in Fig. \ref{teta}. They indicate that the ratio of the nuclear  transparencies is sufficiently sensitive 
to the difference between the interaction cross section of the produced meson with a nucleon and the $\pi N$ cross section.
It is worth emphasizing here that there are many interesting channels in the large angle $\gamma N $ scattering with comparable cross sections: 
several $\pi N$ channels, $K^+\Lambda$ channel, a factor of $\sim 10 $ larger $\rho N$ channel \cite{Anderson:1976ph}.
Other interesting channels include $K^{*+}+\Lambda$ channel which may have even larger cross section than the $K^+\Lambda$ channel  
(if the  pattern observed in the  $\pi$ -  $\rho$- case holds for strange mesons),  non-resonance two-meson production  
channel which may have a very large effective absorption cross section on the nucleon.     
 
In particular, we see that for the $\phi$-production the expected transparency should be $\sim 20\div30\%$ larger than the
transparency for the pion production.
This is because at $t \sim -2$ GeV$^2$ and $p_{\rm lab} \sim 10$ GeV/c the momentum of $\phi$ is $\sim 9$ GeV/c,  
i.e. the $\phi N$ invariant energy, $\sqrt{s_{\phi N}} \sim 4$ GeV, is well above the baryon resonance region.
This makes us more confident that the additive quark model should work in this case. Thus, the expected transparency for the $\eta$-meson production
should be $\sim 10\div15\%$ larger than the pionic transparency. Other limiting case of small transparencies, $\sim 70\div80\%$
of the pionic one, is expected for the two-meson systems (e.g. $\pi\pi, K\bar K$).

Hence, the study of the $A$-dependence of the ratio of transparencies would provide an important test of the interpretation of the data.

     \section{Color transparency effects}
     \label{CT}
The large angle processes are a natural place for looking for the CT effects at the intermediate energies since the unresolved photon can interact with small configurations 
without significant suppression.
 The space-time evolution of the small-size configurations in mesons
was also tested in the studies of the $Q^2$ dependence of the exclusive electroproduction of pions \cite{Clasie:2007aa} 
and $\rho$-mesons \cite{ElFassi:2012nr} at TJNAF.
 
Hence, we can test the sensitivity of the nuclear  transparency to this effect. Similar to the previous studies we use the model discussed in section \ref{formalism}.
We vary the parameter $\Delta M^2$  to estimate the uncertainties of our predictions. In principle, we could have taken the different values of $\Delta M^2$ 
for a meson and a baryon, but it seems not necessary to estimate the magnitude of the expected effect. 
We should emphasize here that we do not modify the basic assumption of the model for $\sigma^{eff}(z=0)$ that the squeezing is  starting at $Q^2\sim \mbox{1 GeV}^2$.
Though this assumption is in line with the TJNAF data on the meson production, it may be too optimistic for the baryon case, cf. discussion in section \ref{Intro} 
and in the review \cite{Dutta:2012ii}.

\begin{figure}
\begin{center}
   \includegraphics[scale = 0.7]{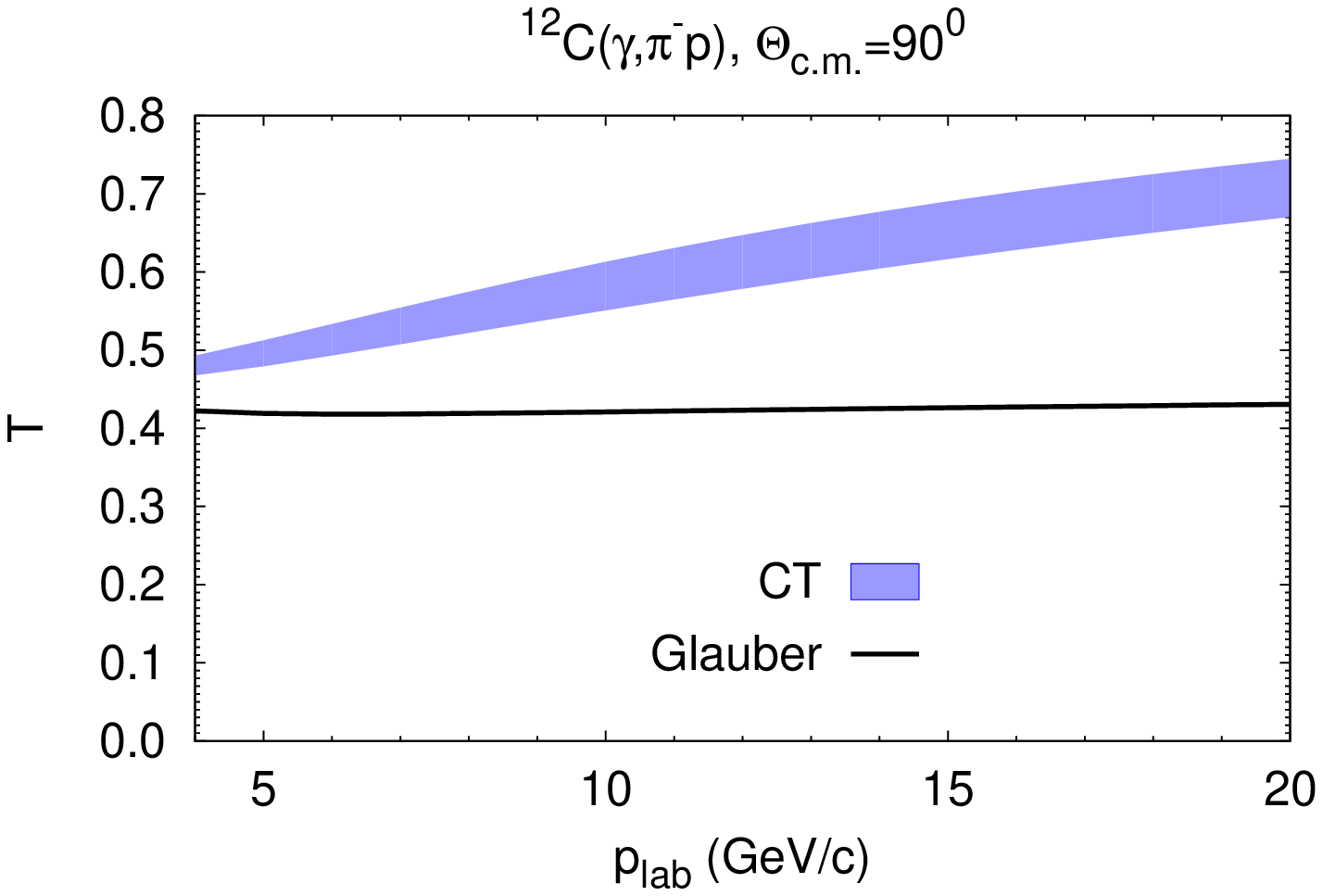}
   \includegraphics[scale = 0.7]{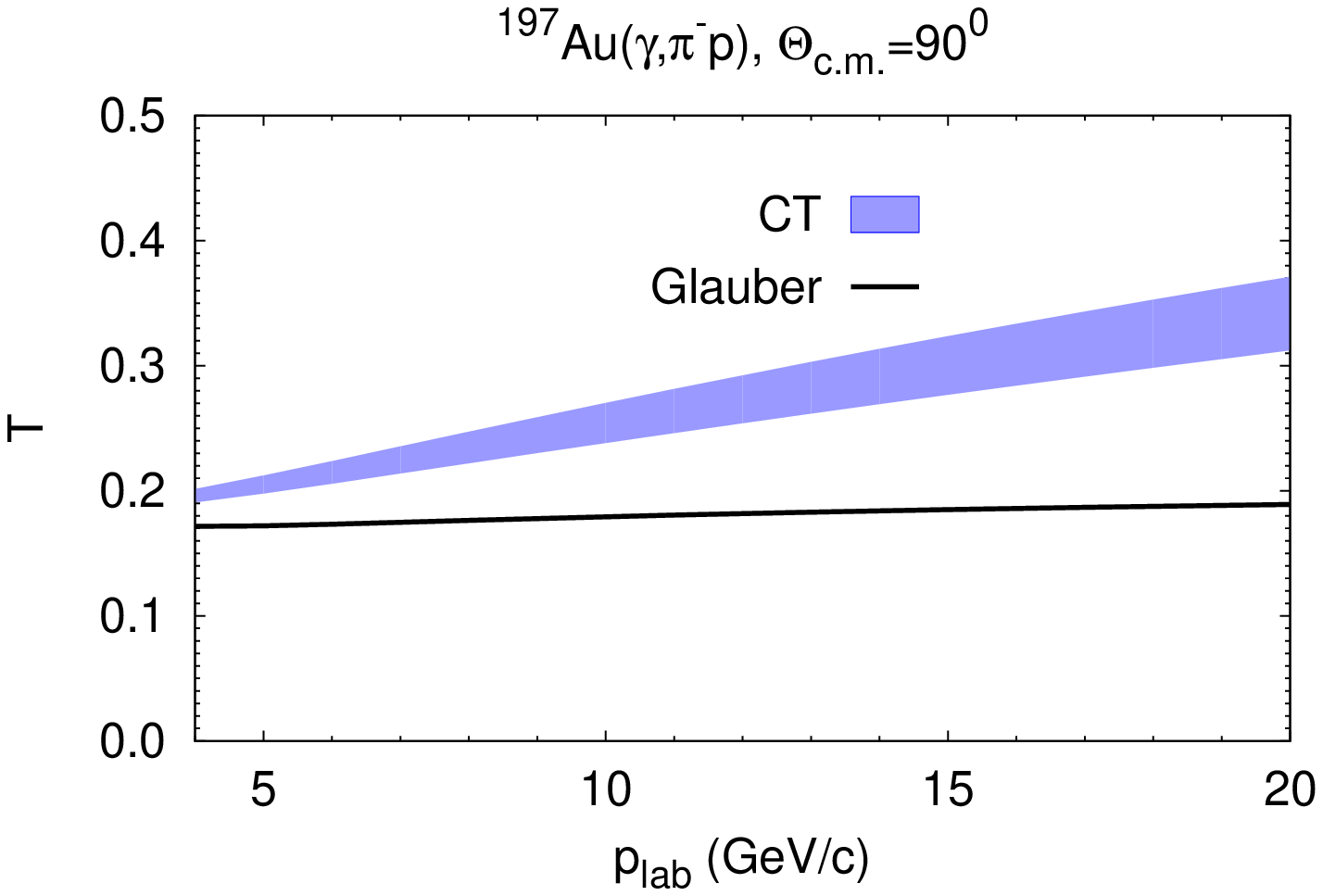}
\end{center}
\caption{\label{fig:T_90deg} The transparency for the $A(\gamma,\pi^-p)$ semiexclusive process for $^{12}$C and $^{197}$Au target nuclei
at $\Theta_{c.m.}=90^o$ vs photon beam momentum. The band and solid line correspond to the quantum diffusion and
Glauber model calculations, respectively. The upper (lower) boundary of the band is given by $\Delta M^2=0.7$ (1.1) GeV$^2$.}
\end{figure}

In Fig.~\ref{fig:T_90deg} we display the nuclear transparency at $\Theta_{c.m.}=90^o$.
The Glauber model calculation produces almost flat $p_{\rm lab}$-dependence of the transparency while the CT effects in the quantum diffusion
model lead to a rather  strong increase of the nuclear transparency with beam momentum.
We see that at $p_{\rm lab}=12$ GeV the transparency is larger by $\sim 50\%$ due to the CT effects 
both for light and heavy nuclear targets.

\begin{figure}
\begin{center}
   \includegraphics[scale = 0.7]{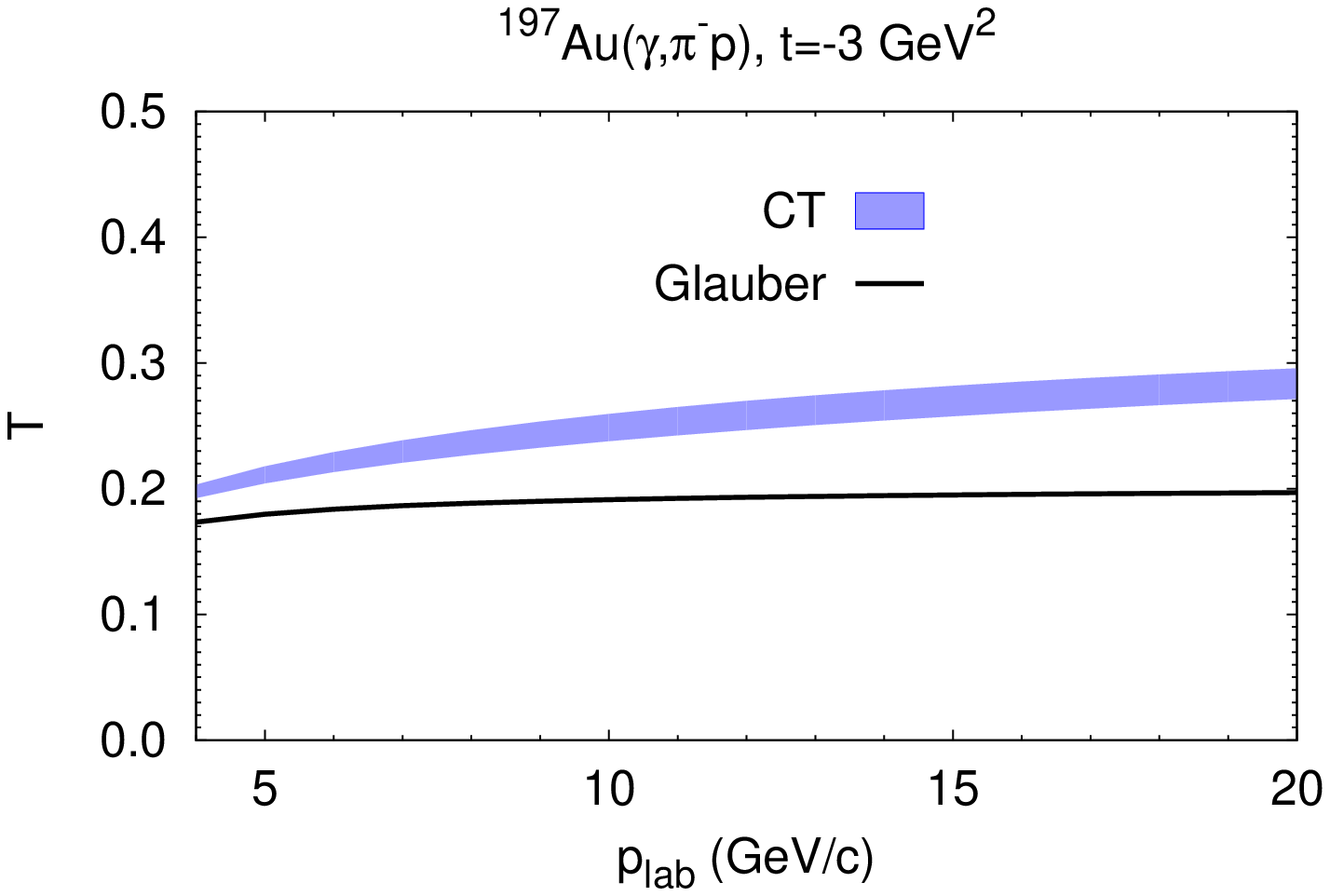}
   \includegraphics[scale = 0.7]{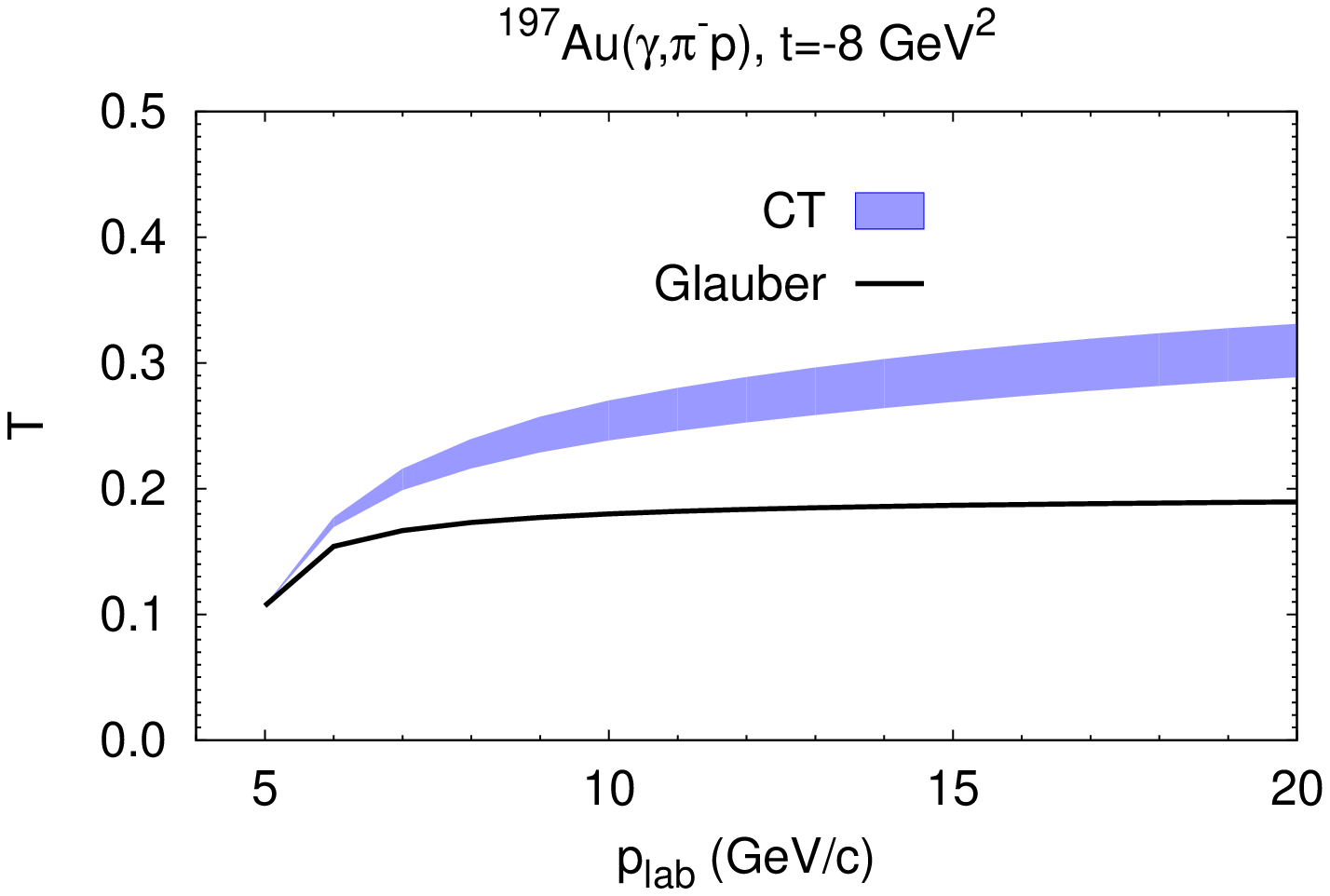}
\end{center}
\caption{\label{fig:T_t} The transparency for the $^{197}$Au$(\gamma,\pi^-p)$ semiexclusive process
at $t=-3$ GeV$^2$ and $t=-8$ GeV$^2$ vs photon beam momentum. 
Notations are the same as in Fig.~\ref{fig:T_90deg}.} 
\end{figure}
In Fig.~\ref{fig:T_t} we present the transparency calculated at fixed $t$. We observe that the choice
of $t=-8$ GeV$^2$ leads to the fall of the transparency at small beam momenta in both calculations, pure Glauber one
and quantum diffusion one. This is due to the increased absorption 
for the forward-backward kinematics at large $|t|$ and small $|u|$. In contrast, in the case of scattering at $\Theta_{c.m.}=90^o$
(Fig.~\ref{fig:T_90deg} lower panel) one of the particles does not propagate through the nucleus in 
peripheral collisions.
The relative CT effect at large beam momenta is somewhat stronger for the case of $\Theta_{c.m.}=90^o$ since in this case the 
coherence lengths of the both outgoing particles grow with beam momentum, while at $t=-8$ GeV$^2$
the momentum of the 
outgoing proton is fixed to $\sim 5$ GeV/c and, thus, its coherence length is also fixed to $\sim 2$ fm.    

\begin{figure}
\begin{center}
   \includegraphics[scale = 0.7]{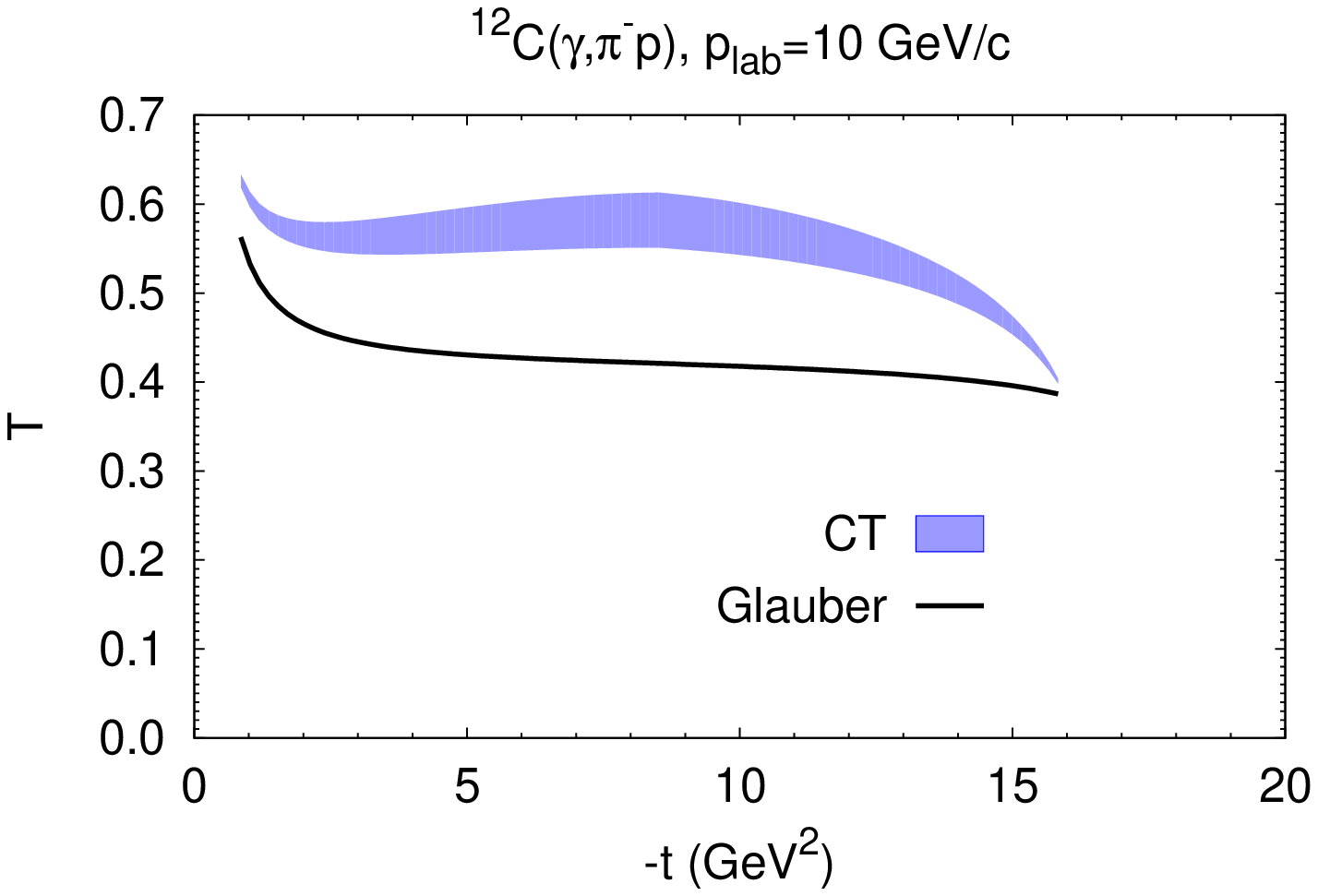}
   \includegraphics[scale = 0.7]{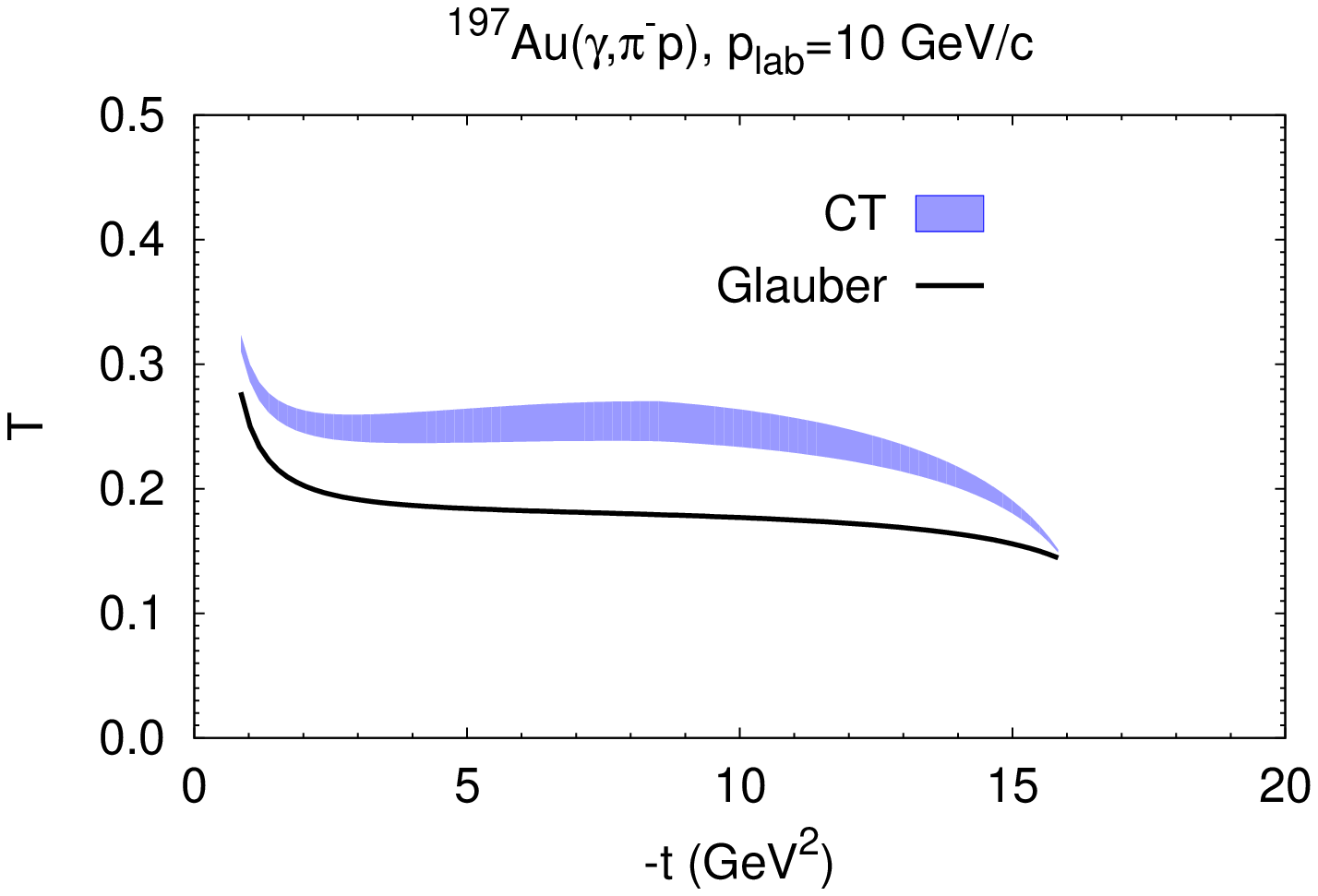}
\end{center}
\caption{\label{fig:T_plab} The transparency for the $A(\gamma,\pi^-p)$ semiexclusive process
on $^{12}$C and $^{197}$Au targets at $p_{\rm lab}=10$ GeV/c vs invariant $-t$. 
Notations are the same as in Fig.~\ref{fig:T_90deg}.} 
\end{figure}
Fig.~\ref{fig:T_plab} displays the transparency $T$ as a function of $-t$ at the fixed beam momentum of 10 GeV/c.
The transparency grows at small $-t$ due to vanishing inelastic $NN$ cross section for a slow proton.
As we see, the effects of CT are strongest at $-t \simeq 9$ GeV$^2$ which corresponds to the scattering at $\Theta_{c.m.}=90^o$.
For the both considered nuclear targets, i.e. for $^{12}$C and $^{197}$Au, the nuclear transparency rises by $\sim 40\%$ 
due to the CT effects.

  \section{Summary and conclusions}
     \label{concl}

To summarize the main results of our calculations:

\begin{itemize}

\item For heavy targets in the case of resolved (i.e. physical) photon the nuclear transparency is expected to be a factor
of $2\div3$ smaller than in the case of unresolved (i.e. bare) photon. This effect is due to the absorption of the hadronic 
component of the physical photon, similar to the high-energy diffractive photoproduction \cite{Bauer:1977iq}.

\item The variation of the inelastic cross section of the outgoing meson - nucleon interaction 
by $\sim 5$ mb results in the change of the nuclear transparency by $\sim 10\%$ 
and, hence, can be observed experimentally.

\item The color transparency leads to the strong rise of the nuclear transparency with
photon beam momentum at fixed $\Theta_{c.m.}$, in contrast to the pure Glauber model calculation.

\end{itemize}

Thus,  we have demonstrated that the studies of photon induced exclusive processes off
nuclei in the kinematics available at TJNAF will provide 
new important insights into the interplay of soft and hard physics in two - body photon-induced reactions.
Moreover, such measurements would make it possible to compare the strengths of the interactions of various
mesons with nucleons. 

These studies would also allow to use the photon-induced reactions as a precision
tool for studying  the short-range nuclear structure including short range
correlations,  non-nucleonic degrees of freedom, etc.

\section*{Acknowledgments}
\label{Ack}
A.L. acknowledges financial support by the Deutsche Forschungsgemeinschaft
(DFG) under Grant No. Le439/9 and the Helmholtz International Center (HIC)
for FAIR.
M.S.'s research was supported by the US Department of Energy Office of Science, 
Office of Nuclear Physics under Award No.  DE-FG02-93ER40771.

\end{document}